\documentclass[times,twocolumn,final]{elsarticle}
\usepackage{prletters}
\biboptions{sort&compress}
\usepackage{amsmath,amssymb,amsfonts}
\usepackage{graphicx}
\usepackage{textcomp}
\usepackage{xcolor}
\usepackage{mathtools}
\usepackage{algorithm}
\usepackage{algorithmicx}
\usepackage{algpseudocode}
\usepackage{amsmath}
\usepackage{times}
\usepackage{mathptmx}
\usepackage{multirow}
\usepackage{subfigure}
\usepackage{url}
\usepackage[justification=centering]{caption}

\journal{Pattern Recognition Letters}
\begin{document}
\clearpage

\ifpreprint
  \setcounter{page}{1}
\else
  \setcounter{page}{1}
\fi

\begin{frontmatter}
\title{Multi-faceted Graph Attention Network for Radar Target Recognition in Heterogeneous Radar Network}
\author[1]{Han Meng}
\ead{menghan@bupt.edu.cn}
\author[1]{Yuexing Peng\corref{cor1}}
\ead{ yxpeng@bupt.edu.cn}
\author[2]{Wei Xiang}
\ead{ w.xiang@latrobe.edu.au}
\author[1]{Xu Pang}
\ead{pengxu_6@bupt.edu.cn}
\author[1]{and Wenbo Wang\corref{cor1}}
\ead{ wbwang@bupt.edu.cn}

\cortext[cor1]{Corresponding Author}
\address[1]{Beijing University of Posts and Telecommunications, Beijing, China}
\address[2]{La Trobe University, Melbourne, VIC, Australia}

\begin{abstract}
      Radar target recognition (RTR), as a key technology of intelligent radar systems, has been well investigated. Accurate RTR at low signal-to-noise ratios (SNRs) still remains an open challenge. Most existing methods are based on a single radar or the homogeneous radar network, which do not fully exploit frequency-dimensional information. In this paper, a two-stream semantic feature fusion model, termed Multi-faceted Graph Attention Network (MF-GAT), is proposed to greatly improve the accuracy in the low SNR region of the heterogeneous radar network. By fusing the features extracted from the source domain and transform domain via a graph attention network model, the MF-GAT model distills higher-level semantic features before classification in a unified framework. Extensive experiments are presented to demonstrate that the proposed model can greatly improve the RTR performance at low SNRs.

\end{abstract}

\end{frontmatter}
\section{Introduction}\label{intro}
Radar target recognition (RTR) is a critical component of modern radar technology, with applications in aviation and other wide-ranging fields. The major concerns of RTR mainly relate to data acquisition, semantic feature discovery and extraction technology. High-resolution signals like high-resolution range profile (HRRP) \cite{feng2017radar,xu2019target}, synthetic aperture radar (SAR) images \cite{cnn1}, and inverse synthetic aperture radar (ISAR) images \cite{isar} present rich information of targets but demand powerful radar. The radar cross section (RCS) signal, which characterizes the scattering shape and the movement pattern of the target, is widely used due to its easy availability and sufficient information for RTR. As a result, the RCS signal is employed for RTR in this study.

RCS signals vary with frequency, illumination directions of the incident wave, scattering coefficient of the surface material, etc \cite{wang2016radar}. It is crucial to reliably extract semantic features from the RCS signal for the purpose of RTR. RCS-based RTR methods can be roughly categorized into the traditional and deep learning based methods. Traditional methods try to detect some special feature parameters to recognize targets. In \cite{lei2011statistical}, 14 statistical features are extracted and a greedy algorithm is employed to discover the best combination of the extracted features for RTR. Besides statistical methods, transform domain methods are also widely applied. In \cite{tang2019radar} 10 statistical features are extracted from the RCS and transform domain via the Mellin transform and then the support vector machine (SVM) and Multi-layer perceptron (MLP) are employed for target recognition. Wang et al. \cite{wang2016radar} extract 5 statistical features via the wavelet transform and establish a set-valued model to represent the correlation between the feature vector and the authenticity of the radar target. There are many methods extracting angular diversity features \cite{chan2013radar,chan2013angular} with physical significance. When features are extracted, several classification criteria are established. In \cite{gokkaya2019novel}, the central moments of the RCS are extracted from different radar targets and then classified via the principal component analysis (PCA) and SVM. In \cite{lee2019radar}, the K-nearest neighbors (KNN) regression method is employed for RTR.

Traditional methods usually demand features engineering and cannot fully extract abstract semantic features. By contrast, deep learning-based methods exhibit superior end-to-end learning and abstract representation capabilities. There exist many popular deep learning models, such as the recurrent neural network (RNN) \cite{xu2019target,ye2021radar, geng2017sar}, generative adversarial network (GAN) \cite{he2019parallel}, and convolutional neural network (CNN) \cite{cnn1}.
In \cite{chen2018convolutional}, a CNN-based RCSNet model is proposed to classify different targets with the same shape. Wengrowski et al. \cite{wengrowski2019deep} simulate the RCS signals of rotating and tumbling targets with unknown motion parameters, and then classify them using a CNN model. The aforementioned methods mainly have two shortcomings, i.e., the dependence on a well-annotated dataset with large volume and variety, and inability to fully exploit inherent semantic features and thus a relatively poor recognition performance at low signal-to-noise ratios (SNRs) when they are based on a single radar or a homogeneous radar array.

A radar array can provide spatial information for RTR. Graph convolutional network (GCN)-based models \cite{kipf2016semi} are investigated recently. It is well known that GCN models depend heavily on the design of the adjacent matrix, which defines the correlation between any two nodes. Existing adjacent matrix design methods are based upon geographical distances \cite{yu2017spatio} or the operating frequencies of the radars \cite{meng}. In the task of RTR, the correlation of the RCS signals at two nodes may does not depend on the geographical distance or operating frequency. In the semantic space, it depends on the observed target. That is, the RCS signals should be highly correlated when the two radars observe the same target regardless of the operating mode or spatial location. In this sense, the semantic similarity in the semantic space should be defined and learned in the GCN model, which facilitates the extension of RTR in the heterogeneous network to extract more information when compared with homogeneous network-based methods.

In this study, a new RTR model, termed the Multi-faced Graph Attention Network (MF-GAT), is developed for heterogeneous radar arrays, where spatially distributed radars may work in different operating modes in terms of operating frequency, bandwidth, pulse width, and pulse repetition interval. The proposed model facilitates the semantic feature extraction in a transform domain by a specifically designed parallel branch, and then fuse features from two branches in the semantic space before deeper abstract feature extraction. Through feature enhancement, the proposed model can greatly improve the RTR performance in the low SNR range.

The rest of the paper is organized as follows. Section~\ref{sec:System_Model} presents the signal and system models. Section~\ref{sec:MF-GAT Model} describes the proposed MFGAT model in detail. Section~\ref{sec:Experient} presents experimental results and analysis, followed by concluding remarks drawn in Section~\ref{sec:Conclusion}.

\section{System Model}\label{sec:System_Model}
A heterogeneous radar network includes $N$ spatially distributed radars with different operating modes and $M$ flight targets. These radars detect targets independently and constitute a heterogeneous radar array.  From the echoes in the form of the RCS, target information such as the scattering shape and movement pattern can be inferred using advanced signal processing. According to electromagnetic theory, the RCS is given by
\begin{equation}\label{2}
g = \mathop {\lim }\limits_{R \to \infty } 4\pi {R^2}{\left| {\frac{{{E_s}}}{{{E_i}}}} \right|^2},\
\end{equation}
where $R$ is the distance from the radar to target, ${{E_s}}$ and ${{E_i}}$ are the scatter field intensity and the incident field intensity, respectively. The RCS is highly influenced by such factors as the carrier frequency and polarization mode of the incident electromagnetic wave, incident angle, surface shape, and scattering coefficient of the target.

During the observation period, the detection range varies with the movement of the target, resulting in fast variations of the received SNR. The received signal can be expressed as  
\begin{equation}\label{3}
x(t) = \alpha (t)g(t) + n(t),\
\end{equation}
where $\alpha (t)$ is the attenuation factor, $g(t)$ is the time varying RCS given by Equation~\ref{2}, and $n(t)$ is the additional white Gaussian noise with zero mean and a variance of $\sigma_n^2$. Then the SNR at the transmitter (TSNR) is defined as $TSNR = \frac{{E\left\{ \left|g \right|^2 \right\}}}{{E\left\{ {{{\left| n \right|}^2}} \right\}}}$, and the SNR at the receiver (RSNR) is defined by $RSNR = E\left\{ {{{\left| {\alpha (t)} \right|}^2}} \right\} \cdot TSNR$.

The heterogeneous radar network is modeled as a undirected graph $G=(X,A)$, where the node feature set $X$ represents the RCS signals at the $N$ radars, and the correlation among the radars is represented by the adjacency matrix $A$ , which demands to learn the semantic similarity between RCS signals of $N$ radars. Based on the heterogenous radar array, the RTR model tries to learn the mapping from the RCS signals to the classification of the observed targets. That is, given RCS signals $X$, the RTR task can be formulated as $Y= F(X)$, where $Y$ denotes the target class label and $F$ denotes the mapping function.

\begin{figure}[t]
\centering
\includegraphics[width=3in]{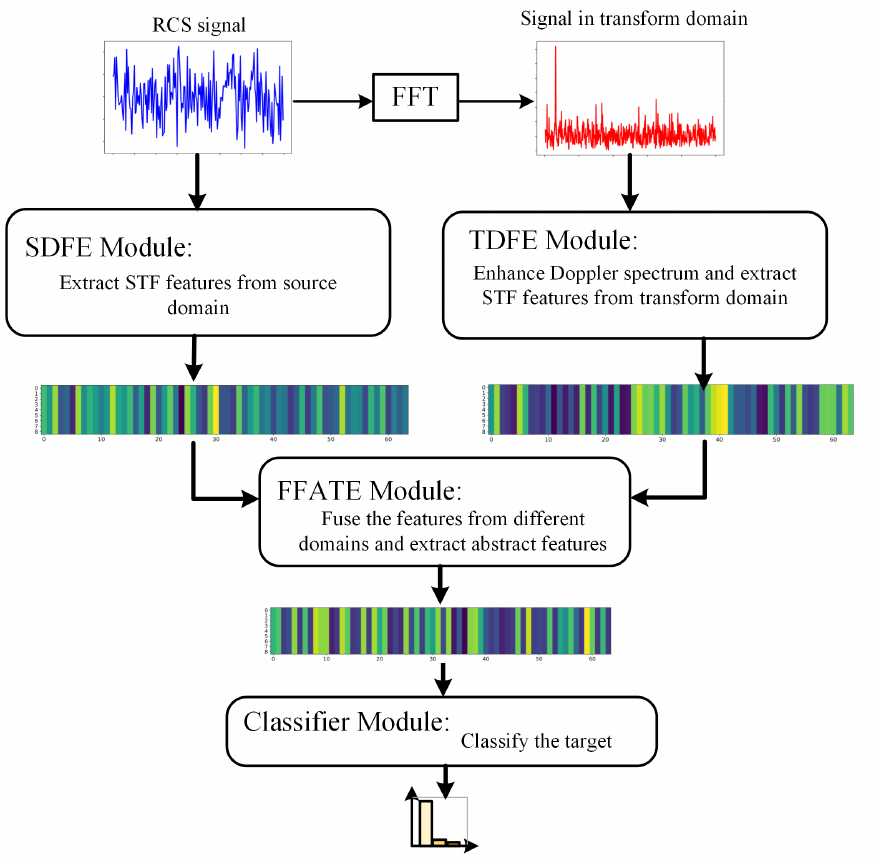}
\caption{The flow chart of the MF-GAT.}
\label{illmodel}

\end{figure}
\begin{figure*}[t]
\centering
\includegraphics[width=\linewidth]{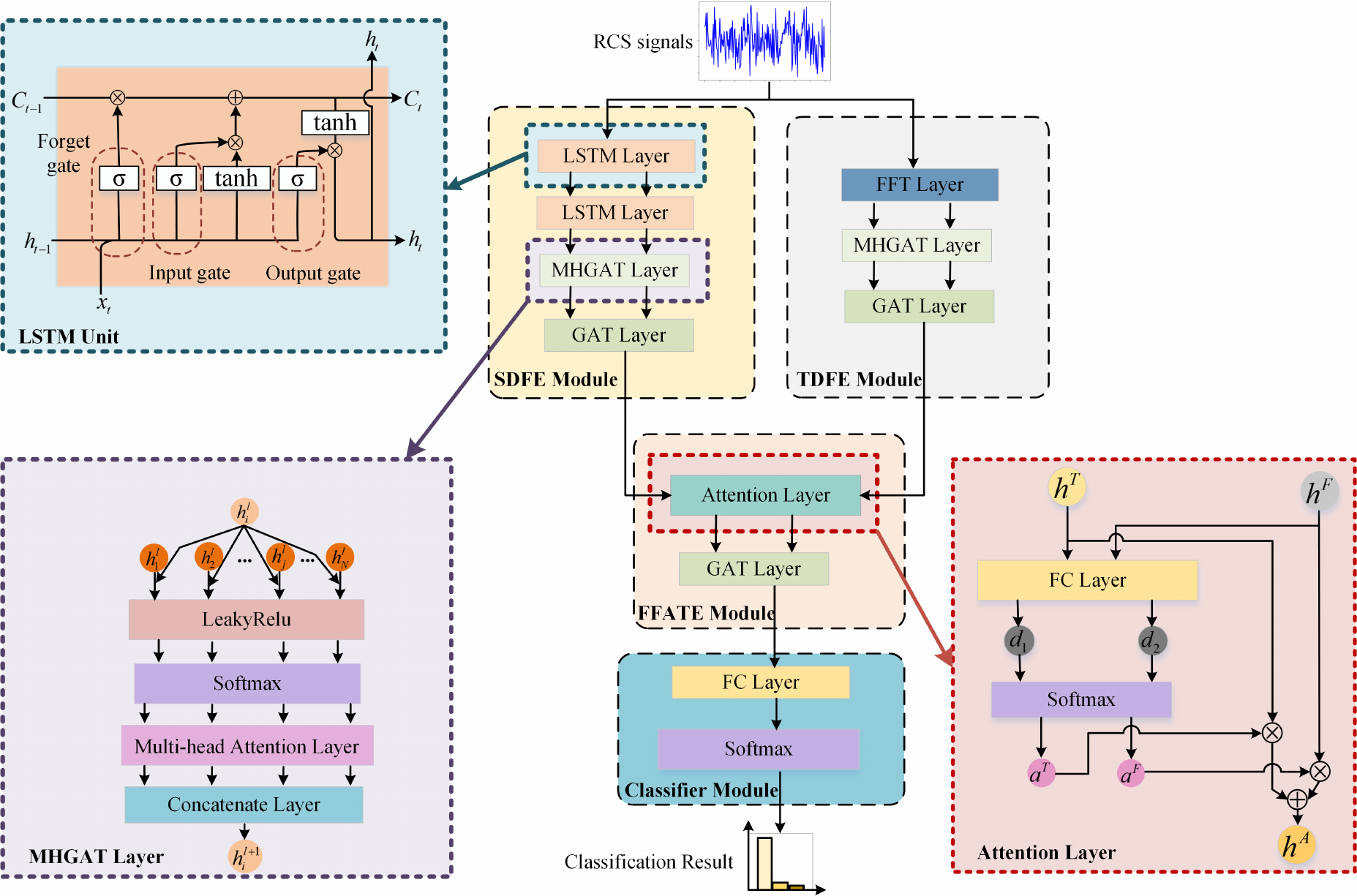}
\caption{The network structure of the MF-GAT.}
\label{model}

\end{figure*}
\section{MF-GAT Model}\label{sec:MF-GAT Model}
In this section, the proposed MF-GAT model is introduced. As depicted in Fig. \ref{illmodel}, the MF-GAT model consists of four modules, including two feature extraction modules, a feature fusion and abstract feature extraction (FFATE) module, and a classifier module. The first feature extraction module extracts semantic features directly from the RCS, which is termed the source domain feature extraction (SDFE) module, while the parallel transform domain feature extraction module termed TDFE extracts semantic features from the transform domain using the fast Fourier transform (FFT). Since the Doppler spectrum is one of the most important feature for moving target recognition, it is necessary to facilitate feature extraction module by presenting signals in frequency domain.

The detailed network structure is shown in Fig.~\ref{model}, in which each  module is detailed subsequently.

\subsection{SDFE Module}\label{AA}

The SDFE module extracts temporal domain features by the long short-term memory (LSTM) network, and then dynamically models the spatial dependencies among the radars by a graph attention network with multi-head attention mechanism (MHGAT) \cite{veli2018graph}, finally fuses multi-domain features using a graph attention network (GAT).
\subsubsection{Temporal Dependency Modeling}
\
\newline
\indent The LSTM network is employed to extract temporal features. An LSTM network is composed of one or more LSTM layers, and each LSTM layer consists of several LSTM units, which inludes three gates, i.e., the forget gate $U$, input gate $I$, and output gate $O$. The operations in an LSTM unit are described below.
\begin{equation}\label{4}
{O_{t,i}} = \sigma ({W_{O,i}}[{h_{t - 1,i}};{x_{t,i}}] + {b_{O,i}}),\
\end{equation}
\begin{equation}\label{3}
{I_{t,i}} = \sigma ({W_{I,i}}[{h_{t - 1,i}};{x_{t,i}}] + {b_{I,i}}),\
\end{equation}
\begin{equation}\label{5}
{U_{t,i}} = \sigma ({W_{U,i}}[{h_{t - 1,i}};{x_{t,i}}] + {b_{U,i}}),\
\end{equation}
\begin{equation}\label{6}
{\widetilde C_{t,i}} = \tanh ({W_{C,i}}[{h_{t - 1,i}};{x_{t,i}}] + {b_{C,i}}),\
\end{equation}
\begin{equation}\label{7}
{C_{t,i}} = {U_{t,i}} \otimes {C_{t - 1,i}} + {I_{t,i}} \otimes {\widetilde C_{t,i}},\
\end{equation}
\begin{equation}\label{8}
{h_{t,i}} = {O_{t,i}} \otimes \tanh ({C_{t,i}}),\
\end{equation}
where $h_{t-1}$ and $h_{t}$ are the input and output of the LSTM unit, respectively; $W_U$, $W_I$, and $W_O$ are the weight matrices to be learnt; $b_U$ , $b_I$, and $b_O$ are the corresponding bias vectors to be learnt; ${\widetilde C_{t,i}}$ and $C$ represent a candidate for cell state and the cell state, respectively.

\subsubsection{Spatial Dependency Modeling}
\
\newline
\indent
Some existing GCN models \cite{yu2017spatio,wu2019graph} consider fixed spatial dependencies according to the geographic distances among nodes in a topology graph. In our model, the semantic similarity of nodes is learned by the GAT, which differs from the GCN in the feature aggregation manner among the neighboring nodes. For a GCN method, the feature aggregation operation returns the standardized sum of the neighbors' features as follows 
\begin{equation}\label{10}
h_i^{l + 1} = \sigma (\sum\limits_{j \in {D_i}} {\frac{1}{{{c_{ij}}}}{W^l}h_j^l} ),\
\end{equation}
where $ \sigma\ $ is the activation function; $D_i$ is the set of radars which are neighbors of the $i$-th radar;  $c_{ij}$ is a normalized constant based on the graph structure; $l$ is the layer index; $W^l$ is a shared weight matrix for feature transformation; and $h_i^l$ is the hidden feature of the $l-$th layer for node $i$.

Based on GCN, GAT learns and weights the semantic features via attention mechanism. The mapping from output $h_i^l$ of the $l$-th layer to the next layer output $h_i^{l+1}$ is shown to be 
\begin{equation}\label{11}
z_i^l = {W^l}h_i^l,\
\end{equation}
\begin{equation}\label{12}
e_{ij}^l = {\rm{LeakyReLU}}({a^l}(z_i^l\parallel z_j^l)),\
\end{equation}
\begin{equation}\label{13}
a_{ij}^l = \frac{{\exp (e_{ij}^l)}}{{\sum\nolimits_{p \in {D_i}} {\exp (e_{ip}^l)} }},\
\end{equation}
\begin{equation}\label{14}
h_i^{l + 1} = \sigma (\sum\nolimits_{j \in {D_i}} {a_{ij}^lz_j^l}).\
\end{equation}
Equation \ref{11} uses a learnable weight matrix $W^l$ to represent a linear combination of output $h_i^l$.  In Equation \ref{12}, a pair-wise un-normalized attention score $e_{ij}^l$ between radars $i$ and $j$ is computed through additive attention, which is performed by concatenating $z_i^l$ and $z_j^l$ via dot product and then weighting the output via a learnable weight vector $a^l$ and the Leaky Rectified Linear Unit (LeakyReLU) activation function \cite{xu2015empirical}. The output is normalized by a softmax activation function in Equation \ref{13}. In Equation \ref{14}, similar to the GCN, the higher layer output $h_i^{l+1}$ is aggregated by the attention scores from neighbors.

To improve the model and stabilize the self-attention learning process, a multi-head attention mechanism \cite{vaswani2017attention} is employed, which allows the model to simultaneously learn the attention scores from various representation sub-spaces. The MHGAT contains independent $K$ multi-head attention mechanisms that execute the GAT convolution operation simultaneously as shown in Fig. \ref{model}. The features are then concatenated to yield the feature representation as follows
\begin{equation}\label{15}
h_i^{l + 1} = \mathop \parallel \limits_{{k} = 1}^{{K}} \sigma (\sum\limits_{j \in {D_i}} {\alpha _{ij}^{{k}}{W^{{k}}}h_j^l} ),\
\end{equation}
where $\mathop \parallel$ represents the concatenation operation.

Both the GAT and MHGAT layers are constructed to extract the spatial correlation among nodes and coherent temporal patterns hidden in the temporal domain. 

\subsection{TDFE Module}

In order to explore features in the Doppler spectrum, the FFT is performed on the RCS signal to facilitate Doppler spectrum feature extraction by the TDFE module, which is formulated as
\begin{equation}\label{9}
F(m) = \sum\limits_{n = 0}^{N - 1} {{g_{real}}(n)} W_N^{mn},m = 0,1,...,N - 1,\
\end{equation}
where ${W_N} = \textrm{exp}(- j2\pi/N)$.

As shown in Fig. \ref{model}, the TDFE module contains the same MHGAT and GAT layers as SDFE. These layers extract the temporal-spatial-frequency features in the transform domain.

%(3) FFATE module dynamically fuses features according to the contribution of the modules mentioned above by an attention mechanism, and extracts high-level feature by a GAT; (4) The classifier module classify the target.

\subsection{FFATE Module and Classifier Module}

As illustrated in Fig. \ref{model}, extracted features from the SDFE and TDFE modules are fused by attention mechanism, and then higher semantic features are distilled through the GAT layer in the FFATE module. The features are fused as follows
\begin{equation}\label{16}
{h^A} = {a^T}{h^T} + {a^F}{h^F},\
\end{equation}
where ${a^T}$ and ${a^F}$ are the attention weights defined by
\begin{equation}\label{17}
{a^S} = {\rm{softmax(tanh(}}{{\rm{W}}^T}{h^T} + {b^T})),\
\end{equation}
\begin{equation}\label{18}
{a^F} = {\rm{softmax(tanh(}}{{\rm{W}}^F}{h^F} + {b^F})).\
\end{equation}

After feature fusion, the semantic features are continually updated by the GAT layer.

In the classifier module, the high-level semantic features are combined through a fully-connected layer, and finally the class label is yielded by the softmax layer.

\section{Experimental Results and Discussions}\label{sec:Experient}
\subsection{Dataset and Data Preprocessing}
The dataset under consideration contains RCS signals received by 9 radar of two types. All the radars independently detect two flight targets. The experimental settings are listed in Table \ref{simulation_settings}. It is assumed that a radar can detect not more than one flight at a time. Then there are three scenarios for each radar, i.e., (1) no target; (2) target A; and (3) target B. The RSNR at each radar varies dynamically due to target movements.

\begin{table}[!h]
	\vspace{0em}
	\caption{Experimental parameters.}
	\begin{center}
	\resizebox{0.99\linewidth}{!}{%
		\begin{tabular}{lc}
			\hline
			\multicolumn{1}{c}{\textbf{Parameters}} & \textbf{Values}                                                                                   \\ \hline
			Number of type-1 radar                      & 5                                                                                                 \\
			Number of type-2 radar           &  4\\
			Bandwidth of radar (MHz)                         & 10                                                                                               \\
			Pulse interval of radar (ms)               & 50                                                                                               \\
			Operating frequency of type-1 radar (GHz)                  & 3.25                                                                                                \\
			Operating frequency of type-1 radar (GHz)       & 2.52                                                                                           \\
			Mass trajectory (km/s)                     &5                                                                                                  \\
			Micro-motion frequency of type-1 aircraft (Hz)                   & 0.64, 2.75                                                                                                \\
			Micro-motion frequency of type-2 aircraft (Hz)       & 1.67, 8.72                                                                           \\ \hline
		\end{tabular}%
		}
	\end{center}
	\label{simulation_settings}
\end{table}
In the dataset, each RCS signal segment includes 105000 samples. The samples are generated by a sliding window method. The size of sliding window is 200 of 10 seconds and the stride is 50. Then 6300 samples are obtained, which are randomly divided into the training, validation and test sets with a ratio of 7:2:1. Finally, all samples are normalized by a zero-mean normalization method.
\subsection{Baseline Models and Experimental Settings}
\begin{table}[!h]
	\vspace{0em}
	\caption{Experiment hyper-parameters.}
	\begin{center}
	\resizebox{0.8\linewidth}{!}{%	
		\begin{tabular}{lc}
			\hline
			\multicolumn{1}{c}{\textbf{Hyper-parameters}} & \textbf{Values}                                                                                   \\ \hline
			 Input units of LSTM       & 200                          \\
			 Hidden units of LSTM       & 128                              \\
			Number of training epochs                      & 100                                             \\
			Initial Learning rate           &  0.0005\\
			Dropout probability                          & 0.6                                            \\
			Optimizer               & Adam                                                          \\
			Batch size       & 32                                  \\
			Loss function            &CrossEntropyLoss                                                                  \\
			Attention heads in MHGAT                   & 8                                   \\
			
			 \hline
		\end{tabular}%
		}
	\end{center}
	\label{Hyperparameters}
\end{table}
    All the comparative models are evaluated by Pytorch on NVIDIA GeForce GTX 3090. The experimental hyper-parameter settings are listed in Table~\ref{Hyperparameters}. The employed performance evaluation metric is accuracy, which is define by
    	\begin{equation}\label{51}
    		Accuracy = \frac{{TP + TN}}{{TP + TN + FP + FN}},\
    	\end{equation}
   where $TP$ and $TN$ are the correct numbers of positive samples and negative samples, respectively. $FP$ and $FN$ are the incorrect numbers of positive samples and negative samples, respectively. .	

The compared baseline models include the following: 
\begin{itemize}
\item Temporal dimension model: LSTM \cite{sehgal2019automatic} \item  Frequency dimension model: FFT-based CLEAN \cite{choi2014bistatic} \item  Temporal-spatial (TS) dimension model: STGCN \cite{yu2017spatio},\item  Temporal-spatial-frequency (TSF) dimension algorithm: STFGACN \cite{meng}.
\end{itemize}

\begin{figure}[t]
   	\centerline{\includegraphics[width=\linewidth]{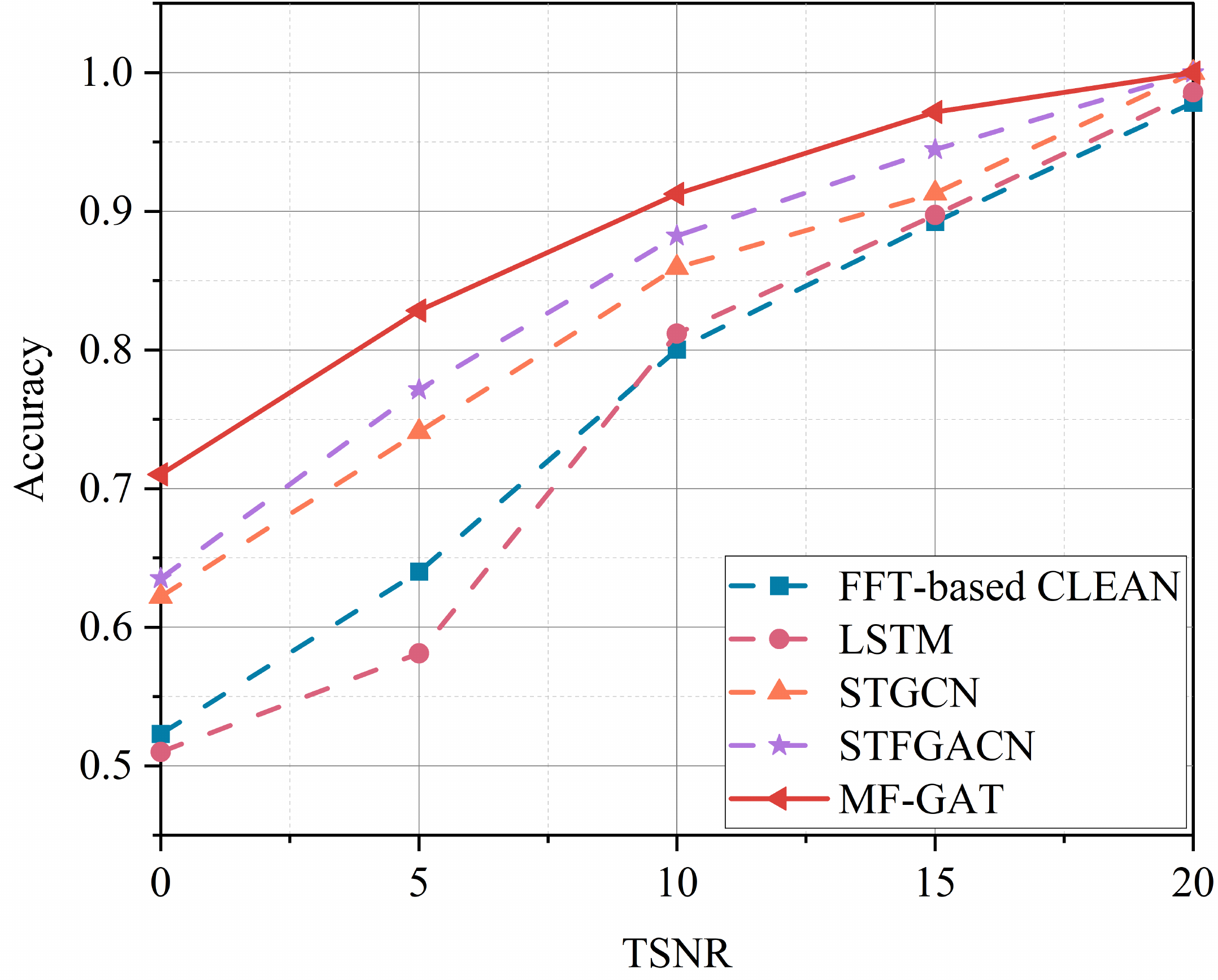}}
   	\caption{Accuracy performance comparison.}
   	\vspace{-0.3cm}
   	\label{result}
\end{figure}

\subsection{Main Results}
The experimental results on accuracy versus the TSNR in dB are shown in Fig. \ref{result}. As stated before, the RSNR varies independently among radars due to the distance changes continuously caused by target movements. As a result, the TSNR is considered instead for fair comparison. In our experiments, the RSNR values are 7.5 dB below than the TSNR on average.

It can be observed from Fig.\ref{result}:
\begin{itemize}
\item The proposed MF-GAT model outperforms all the reference models in the entire RSNR region, especially at low SNRs. . Specifically, at the low TSNR of 0 dB, the MT-GAT achieves 71.0$\%$ accuracy, which is $20.0\%$, $18.5\%$, $8.8\%$, $7.5\%$ better than the FFT-based CLEAN, LSTM, STGCN, STFGACN models, respectively.

\item GCN-based multi-dimension models, including STGCN, STFGACN and MF-GAT, achieve better and more stable performance than single dimension models like FFT-based CLEAN and LSTM. This validates the fact that extracting features from multiple dimensions is able to improve accuracy, and the improvement increases with the dimension of domains. This is further verified by the results that STFGACN and MF-GAT behave better than STGCN.

\item The gain in accuracy achieved by the proposed model increases with the decrease of SNR. Compared to the STFGACN model that extracts semantic features in the same temporal-spatial-frequency domain, the proposed MT-GAT model benefits from the extra parallel branch to extract the most critical Doppler spectrum feature in the transform domain, which results in the an evident improvement in RTR at low SNRs.

\item By achieving the accuracy of 0.85, the MT-GAT achieves 3 dB and over 5 dB SNR gains over the GCN-based models and other models, respectively. The SNR gains can be translated into a longer detection range, which is vital for RTR.

\end{itemize}

\subsection{Ablation Studies}
Ablation experiments are carried out to shed more light on the performance gains incurred by the functional modules in the MF-GAT model, i.e., the parallel branch and the multi-faceted fusing module. Three variants of the MF-GAT model are evaluated as follows:

\begin{itemize}
\item Baseline model, denoted by SDFE, only contains the SDFE and classifier modules, which extracts semantic features from the RCS signal directly.

\item Baseline + TDFE, denoted by STDFE, contains both the TDFE and SDFE modules, but the FFATE module is replaced by the concatenation operation, which means features from the TDFE and SDFE modules are concatenated via channel fusion.

\item Baseline + TDFE + FFATE, denoted by MF-GAT, is the complete model.
\end{itemize}
\begin{figure}[htbp]
\centering
\subfigure[TSNR=0 dB]
{
	\centering
	\includegraphics[width=0.23\textwidth]{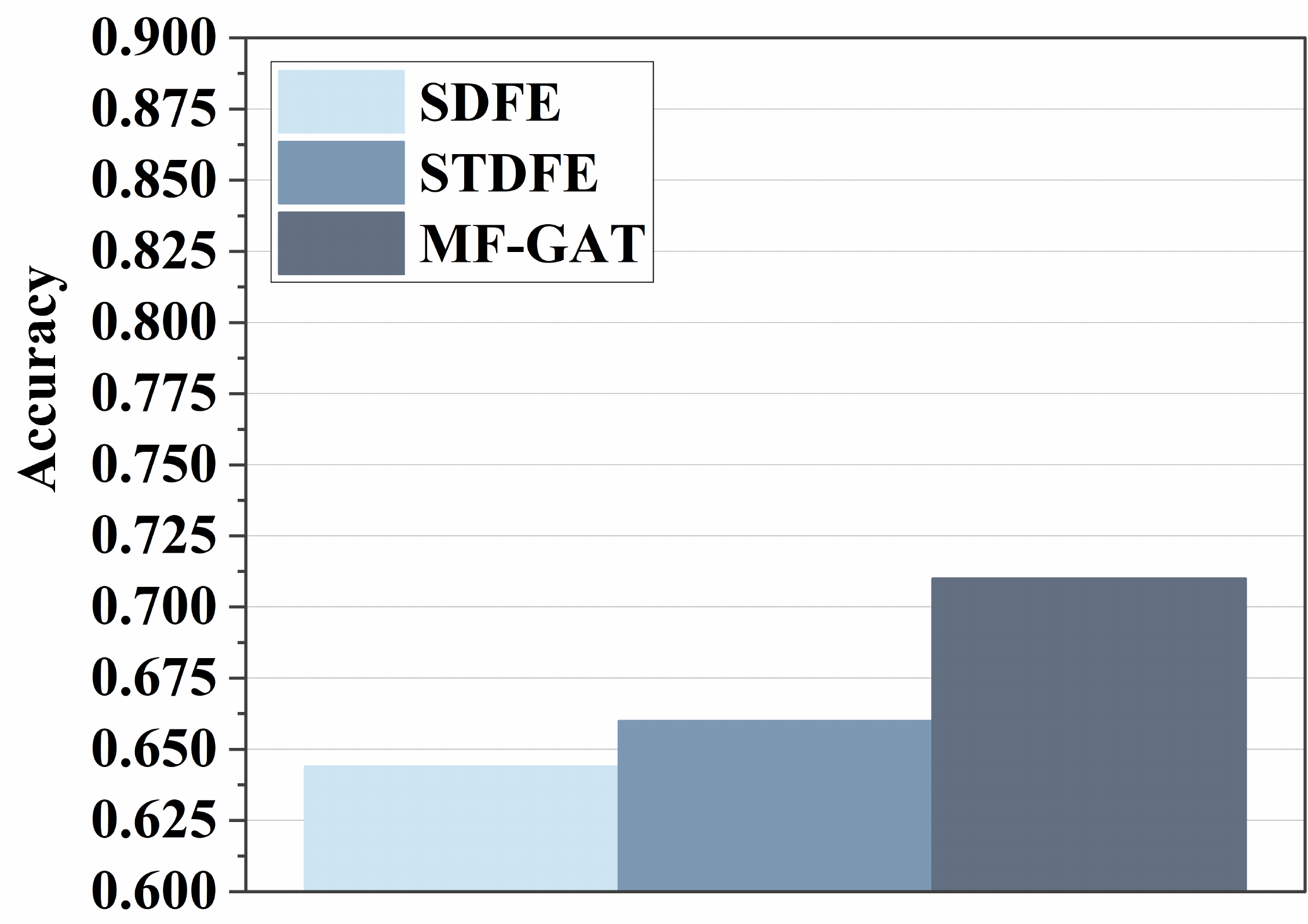}}
\subfigure[TSNR=5 dB]
{
	\centering
	\includegraphics[width=0.23\textwidth]{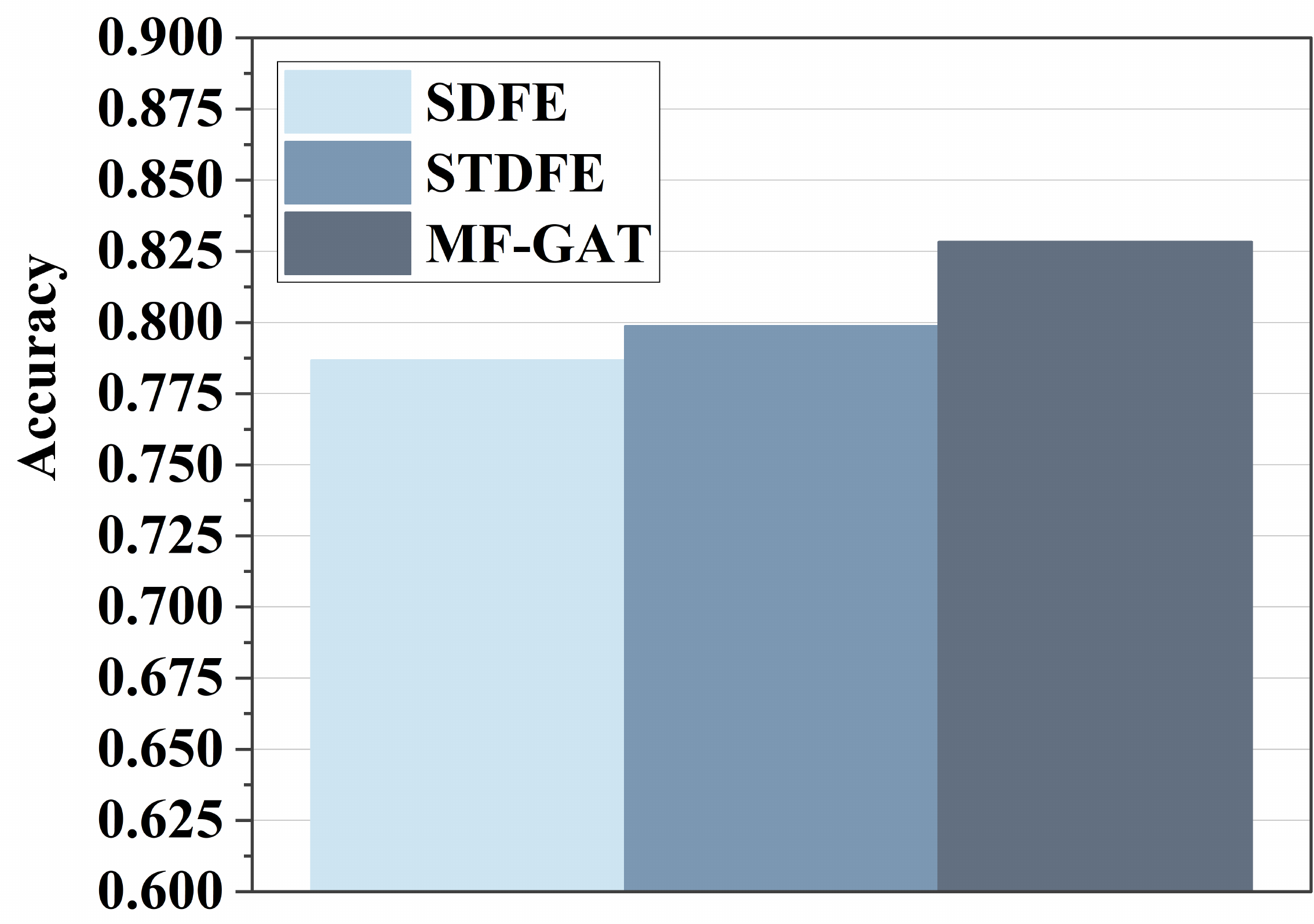}}
\caption{Ablation studies on different components.}
\label{ablation}
\end{figure}

Fig. \ref{ablation} presents the ablation results on the accuracy of RTR in two TSNR settings. As can be observed from the figure, the TDFE and FFATE modules can provide 1.2\% and 2.9\% accuracy enhancement at the TSNR of 5 dB, respectively, and the performance gains increase to 1.6\% and 5\% at the TSNR of 0 dB. Obviously, at the low SNRs, the extra parallel branch and the semantic feature fusion by the FFATE module are able to make RTR more reliable and more accurate.

\section{Conclusion}\label{sec:Conclusion}
In this paper, the MF-GAT modelwas proposed with the objective of greatly improving the accuracy of the RTR task in low SRN on a heterogenous radar arrays at low SNRs. The proposed MF-GAT benefits from both the information dimension extension to temporal-spatial-frequency dimension and the extraction of Doppler spectrum features in the transform domain. Both comparative experimental and ablation results were presented to verify that the proposed MF-GAT model can reliably extract semantic features at low SNRs such that more accurate RTR is achieved.
\section*{Acknowledgments}
This work is supported by NSFC under grant 62071063.
\bibliographystyle{model1-num-names}
\bibliography{GATref}

\begin{thebibliography}{25}
\expandafter\ifx\csname natexlab\endcsname\relax\def\natexlab#1{#1}\fi
\providecommand{\url}[1]{\texttt{#1}}
\providecommand{\href}[2]{#2}
\providecommand{\path}[1]{#1}
\providecommand{\DOIprefix}{doi:}
\providecommand{\ArXivprefix}{arXiv:}
\providecommand{\URLprefix}{URL: }
\providecommand{\Pubmedprefix}{pmid:}
\providecommand{\doi}[1]{\href{http://dx.doi.org/#1}{\path{#1}}}
\providecommand{\Pubmed}[1]{\href{pmid:#1}{\path{#1}}}
\providecommand{\bibinfo}[2]{#2}
\ifx\xfnm\relax \def\xfnm[#1]{\unskip,\space#1}\fi
%Type = Article
\bibitem[{Feng et~al.(2017)Feng, Chen, and Liu}]{feng2017radar}
\bibinfo{author}{B.~Feng}, \bibinfo{author}{B.~Chen}, \bibinfo{author}{H.~Liu},
\newblock \bibinfo{title}{Radar {HRRP} target recognition with deep networks},
\newblock \bibinfo{journal}{Pattern Recognition} \bibinfo{volume}{61}
  (\bibinfo{year}{2017}) \bibinfo{pages}{379--393}.
%Type = Article
\bibitem[{Xu et~al.(2019)Xu, Chen, Wan, Liu, and Jin}]{xu2019target}
\bibinfo{author}{B.~Xu}, \bibinfo{author}{B.~Chen}, \bibinfo{author}{J.~Wan},
  \bibinfo{author}{H.~Liu}, \bibinfo{author}{L.~Jin},
\newblock \bibinfo{title}{Target-aware recurrent attentional network for radar
  {HRRP} target recognition},
\newblock \bibinfo{journal}{Signal Processing} \bibinfo{volume}{155}
  (\bibinfo{year}{2019}) \bibinfo{pages}{268--280}.
%Type = Article
\bibitem[{Xue et~al.(2020)Xue, Bai, and Zhou}]{cnn1}
\bibinfo{author}{R.~Xue}, \bibinfo{author}{X.~Bai}, \bibinfo{author}{F.~Zhou},
\newblock \bibinfo{title}{Spatial--temporal ensemble convolution for sequence
  {SAR} target classification},
\newblock \bibinfo{journal}{IEEE Transactions on Geoscience and Remote Sensing}
  \bibinfo{volume}{59} (\bibinfo{year}{2020}) \bibinfo{pages}{1250--1262}.
%Type = Article
\bibitem[{Luo et~al.(2009)Luo, Zhang, Qiu, Liang, and Li}]{isar}
\bibinfo{author}{Y.~Luo}, \bibinfo{author}{Q.~Zhang}, \bibinfo{author}{C.-w.
  Qiu}, \bibinfo{author}{X.-j. Liang}, \bibinfo{author}{K.-m. Li},
\newblock \bibinfo{title}{Micro-doppler effect analysis and feature extraction
  in {ISAR} imaging with stepped-frequency chirp signals},
\newblock \bibinfo{journal}{IEEE Transactions on Geoscience and Remote Sensing}
  \bibinfo{volume}{48} (\bibinfo{year}{2009}) \bibinfo{pages}{2087--2098}.
%Type = Article
\bibitem[{Wang et~al.(2016)Wang, Bi, Zhao, and Xue}]{wang2016radar}
\bibinfo{author}{T.~Wang}, \bibinfo{author}{W.~Bi}, \bibinfo{author}{Y.~Zhao},
  \bibinfo{author}{W.~Xue},
\newblock \bibinfo{title}{Radar target recognition algorithm based on {RCS}
  observation sequence—set-valued identification method},
\newblock \bibinfo{journal}{Journal of Systems Science and Complexity}
  \bibinfo{volume}{29} (\bibinfo{year}{2016}) \bibinfo{pages}{573--588}.
%Type = Inproceedings
\bibitem[{Lei et~al.(2011)Lei, Fu, Wang, and Gao}]{lei2011statistical}
\bibinfo{author}{X.~Lei}, \bibinfo{author}{X.~Fu}, \bibinfo{author}{C.~Wang},
  \bibinfo{author}{M.~Gao},
\newblock \bibinfo{title}{Statistical feature selection of narrowband {RCS}
  sequence based on greedy algorithm},
\newblock in: \bibinfo{booktitle}{Proceedings of 2011 IEEE CIE International
  Conference on Radar}, volume~\bibinfo{volume}{2},
  \bibinfo{organization}{IEEE}, \bibinfo{year}{2011}, pp.
  \bibinfo{pages}{1664--1667}.
%Type = Inproceedings
\bibitem[{Tang et~al.(2019)Tang, Yu, Wei, and Tong}]{tang2019radar}
\bibinfo{author}{W.~Tang}, \bibinfo{author}{L.~Yu}, \bibinfo{author}{Y.~Wei},
  \bibinfo{author}{P.~Tong},
\newblock \bibinfo{title}{Radar target recognition of ballistic missile in
  complex scene},
\newblock in: \bibinfo{booktitle}{2019 IEEE International Conference on Signal,
  Information and Data Processing (ICSIDP)}, \bibinfo{organization}{IEEE},
  \bibinfo{year}{2019}, pp. \bibinfo{pages}{1--6}.
%Type = Article
\bibitem[{Chan and Lee(2013{\natexlab{a}})}]{chan2013radar}
\bibinfo{author}{S.-C. Chan}, \bibinfo{author}{K.-C. Lee},
\newblock \bibinfo{title}{Radar target recognition by {MSD} algorithms on
  angular-diversity {RCS}},
\newblock \bibinfo{journal}{IEEE Antennas and Wireless Propagation Letters}
  \bibinfo{volume}{12} (\bibinfo{year}{2013}{\natexlab{a}})
  \bibinfo{pages}{937--940}.
%Type = Article
\bibitem[{Chan and Lee(2013{\natexlab{b}})}]{chan2013angular}
\bibinfo{author}{S.-C. Chan}, \bibinfo{author}{K.-C. Lee},
\newblock \bibinfo{title}{Angular-diversity target recognition by kernel
  scatter-difference based discriminant analysis on {RCS}},
\newblock \bibinfo{journal}{International Journal of Applied Electromagnetics
  and Mechanics} \bibinfo{volume}{42} (\bibinfo{year}{2013}{\natexlab{b}})
  \bibinfo{pages}{409--420}.
%Type = Inproceedings
\bibitem[{G{\"o}kkaya and G{\"u}nel(2019)}]{gokkaya2019novel}
\bibinfo{author}{E.~G{\"o}kkaya}, \bibinfo{author}{T.~G{\"u}nel},
\newblock \bibinfo{title}{A novel hybrid approach for radar target
  classification based on {SVM} and central moments with {PCA} using rcs},
\newblock in: \bibinfo{booktitle}{2019 11th International Conference on
  Electrical and Electronics Engineering (ELECO)},
  \bibinfo{organization}{IEEE}, \bibinfo{year}{2019}, pp.
  \bibinfo{pages}{575--579}.
%Type = Article
\bibitem[{Lee(2019)}]{lee2019radar}
\bibinfo{author}{K.-C. Lee},
\newblock \bibinfo{title}{Radar target recognition by machine learning of
  k-nearest neighbors regression on angular diversity {RCS}},
\newblock \bibinfo{journal}{The Applied Computational Electromagnetics Society
  Journal (ACES)}  (\bibinfo{year}{2019}) \bibinfo{pages}{75--81}.
%Type = Article
\bibitem[{Ye et~al.(2021)Ye, Hu, Yan, Meng, Zhu, and Xu}]{ye2021radar}
\bibinfo{author}{L.~Ye}, \bibinfo{author}{S.~Hu}, \bibinfo{author}{T.~Yan},
  \bibinfo{author}{X.~Meng}, \bibinfo{author}{M.~Zhu}, \bibinfo{author}{R.~Xu},
\newblock \bibinfo{title}{Radar target shape recognition using a gated
  recurrent unit based on {RCS} time series' statistical features by sliding
  window segmentation},
\newblock \bibinfo{journal}{IET Radar, Sonar \& Navigation}
  \bibinfo{volume}{15} (\bibinfo{year}{2021}) \bibinfo{pages}{1715--1726}.
%Type = Article
\bibitem[{Geng et~al.(2017)Geng, Wang, Fan, and Ma}]{geng2017sar}
\bibinfo{author}{J.~Geng}, \bibinfo{author}{H.~Wang}, \bibinfo{author}{J.~Fan},
  \bibinfo{author}{X.~Ma},
\newblock \bibinfo{title}{{SAR} image classification via deep recurrent
  encoding neural networks},
\newblock \bibinfo{journal}{IEEE Transactions on Geoscience and Remote Sensing}
  \bibinfo{volume}{56} (\bibinfo{year}{2017}) \bibinfo{pages}{2255--2269}.
%Type = Article
\bibitem[{He et~al.(2019)He, Xiong, Zhang, and Liao}]{he2019parallel}
\bibinfo{author}{C.~He}, \bibinfo{author}{D.~Xiong},
  \bibinfo{author}{Q.~Zhang}, \bibinfo{author}{M.~Liao},
\newblock \bibinfo{title}{Parallel connected generative adversarial network
  with quadratic operation for {SAR} image generation and application for
  classification},
\newblock \bibinfo{journal}{Sensors} \bibinfo{volume}{19}
  (\bibinfo{year}{2019}) \bibinfo{pages}{871}.
%Type = Article
\bibitem[{Chen et~al.(2018)Chen, Xu, and Chen}]{chen2018convolutional}
\bibinfo{author}{J.~Chen}, \bibinfo{author}{S.~Xu}, \bibinfo{author}{Z.~Chen},
\newblock \bibinfo{title}{Convolutional neural network for classifying space
  target of the same shape by using {RCS} time series},
\newblock \bibinfo{journal}{IET Radar, Sonar \& Navigation}
  \bibinfo{volume}{12} (\bibinfo{year}{2018}) \bibinfo{pages}{1268--1275}.
%Type = Article
\bibitem[{Wengrowski et~al.(2019)Wengrowski, Purri, Dana, and
  Huston}]{wengrowski2019deep}
\bibinfo{author}{E.~Wengrowski}, \bibinfo{author}{M.~Purri},
  \bibinfo{author}{K.~Dana}, \bibinfo{author}{A.~Huston},
\newblock \bibinfo{title}{Deep {CNN}s as a method to classify rotating objects
  based on monostatic {RCS}},
\newblock \bibinfo{journal}{IET Radar, Sonar \& Navigation}
  \bibinfo{volume}{13} (\bibinfo{year}{2019}) \bibinfo{pages}{1092--1100}.
%Type = Inproceedings
\bibitem[{Kipf and Welling(2017)}]{kipf2016semi}
\bibinfo{author}{T.~N. Kipf}, \bibinfo{author}{M.~Welling},
\newblock \bibinfo{title}{Semi-supervised classification with graph
  convolutional networks},
\newblock in: \bibinfo{booktitle}{Proceedings of International Conference on
  Learning Representations}, \bibinfo{year}{2017}.
%Type = Inproceedings
\bibitem[{Yu et~al.(2018)Yu, Yin, and Zhu}]{yu2017spatio}
\bibinfo{author}{B.~Yu}, \bibinfo{author}{H.~Yin}, \bibinfo{author}{Z.~Zhu},
\newblock \bibinfo{title}{Spatio-temporal graph convolutional networks: {A}
  deep learning framework for traffic forecasting},
\newblock in: \bibinfo{booktitle}{Proceedings of International Joint Conference
  on Artificial Intelligence}, \bibinfo{year}{2018}, pp.
  \bibinfo{pages}{3634--3640}.
%Type = Article
\bibitem[{Meng et~al.(2022)Meng, Peng, Wang, Cheng, Li, and Xiang}]{meng}
\bibinfo{author}{H.~Meng}, \bibinfo{author}{Y.~Peng},
  \bibinfo{author}{W.~Wang}, \bibinfo{author}{P.~Cheng},
  \bibinfo{author}{Y.~Li}, \bibinfo{author}{W.~Xiang},
\newblock \bibinfo{title}{Spatio-temporal-frequency graph attention
  convolutional network for aircraft recognition based on heterogeneous radar
  network},
\newblock \bibinfo{journal}{IEEE Transactions on Aerospace and Electronic
  Systems}  (\bibinfo{year}{2022}).
%Type = Inproceedings
\bibitem[{Veličković et~al.(2018)Veličković, Cucurull, Casanova, Romero,
  Liò, and Bengio}]{veli2018graph}
\bibinfo{author}{P.~Veličković}, \bibinfo{author}{G.~Cucurull},
  \bibinfo{author}{A.~Casanova}, \bibinfo{author}{A.~Romero},
  \bibinfo{author}{P.~Liò}, \bibinfo{author}{Y.~Bengio},
\newblock \bibinfo{title}{Graph attention networks},
\newblock in: \bibinfo{booktitle}{Proceedings of International Conference on
  Learning Representations}, \bibinfo{year}{2018}.
%Type = Inproceedings
\bibitem[{Wu et~al.(2019)Wu, Pan, Long, Jiang, and Zhang}]{wu2019graph}
\bibinfo{author}{Z.~Wu}, \bibinfo{author}{S.~Pan}, \bibinfo{author}{G.~Long},
  \bibinfo{author}{J.~Jiang}, \bibinfo{author}{C.~Zhang},
\newblock \bibinfo{title}{Graph wavenet for deep spatial-temporal graph
  modeling},
\newblock in: \bibinfo{booktitle}{Proceedings of the 28th International Joint
  Conference on Artificial Intelligence}, \bibinfo{year}{2019}, pp.
  \bibinfo{pages}{1907--1913}.
%Type = Inproceedings
\bibitem[{Xu et~al.(2015)Xu, Wang, Chen, and Li}]{xu2015empirical}
\bibinfo{author}{B.~Xu}, \bibinfo{author}{N.~Wang}, \bibinfo{author}{T.~Chen},
  \bibinfo{author}{M.~Li},
\newblock \bibinfo{title}{Empirical evaluation of rectified activations in
  convolutional network},
\newblock in: \bibinfo{booktitle}{Proceedings of the 32th International
  Conference on Machine Learning: Deep Learning Workshop},
  \bibinfo{year}{2015}.
%Type = Article
\bibitem[{Vaswani et~al.(2017)Vaswani, Shazeer, Parmar, Uszkoreit, Jones,
  Gomez, Kaiser, and Polosukhin}]{vaswani2017attention}
\bibinfo{author}{A.~Vaswani}, \bibinfo{author}{N.~Shazeer},
  \bibinfo{author}{N.~Parmar}, \bibinfo{author}{J.~Uszkoreit},
  \bibinfo{author}{L.~Jones}, \bibinfo{author}{A.~N. Gomez},
  \bibinfo{author}{{\L}.~Kaiser}, \bibinfo{author}{I.~Polosukhin},
\newblock \bibinfo{title}{Attention is all you need},
\newblock \bibinfo{journal}{Advances in neural information processing systems}
  \bibinfo{volume}{30} (\bibinfo{year}{2017}).
%Type = Inproceedings
\bibitem[{Sehgal et~al.(2019)Sehgal, Shekhawat, and Jana}]{sehgal2019automatic}
\bibinfo{author}{B.~Sehgal}, \bibinfo{author}{H.~S. Shekhawat},
  \bibinfo{author}{S.~K. Jana},
\newblock \bibinfo{title}{Automatic target recognition using recurrent neural
  networks},
\newblock in: \bibinfo{booktitle}{2019 International Conference on Range
  Technology (ICORT)}, \bibinfo{organization}{IEEE}, \bibinfo{year}{2019}, pp.
  \bibinfo{pages}{1--5}.
%Type = Inproceedings
\bibitem[{Choi and Lee(2014)}]{choi2014bistatic}
\bibinfo{author}{I.-S. Choi}, \bibinfo{author}{S.-J. Lee},
\newblock \bibinfo{title}{Bistatic radar target identification using
  {FFT}-based {CLEAN}},
\newblock in: \bibinfo{booktitle}{2014 IEEE Geoscience and Remote Sensing
  Symposium}, \bibinfo{organization}{IEEE}, \bibinfo{year}{2014}, pp.
  \bibinfo{pages}{1825--1828}.

\end{thebibliography}

\end{document}